\documentstyle[11pt,fleqn]{article}
\input{psfig.tex}
\begin{document}
\title{Stability of Quantum Fluids : Wavy Interface Effect}
\author{A. Kwang-Hua CHU \cite{Chu:2004}} 
\date{P.O. Box 39, Tou-Di-Ban, Road XiHong, Urumqi 830000, PR China}
\maketitle
\begin{abstract}
A numerical investigation for the stability of the incompressible
slip flow of normal quantum fluids (above the critical phase
transition temperature) inside a microslab where surface acoustic
waves propagate along the walls is presented. Governing equations
and associated slip velocity and wavy interface boundary
conditions for the flow of normal fluids confined between elastic
wavy interfaces are obtained. The numerical approach is an
extension (with a complex matrix pre-conditioning) of the spectral
method. We found that the critical Reynolds number ($Re_{cr}$ or
the critical velocity) decreases significantly once the slip
velocity and wavy interface effects are present and the latter is
dominated ($Re_{cr}$ mainly depends on the wavy interfaces).
\newline

\end{abstract}
\doublerulesep=6.5mm \baselineskip=6.5mm
\bibliographystyle{plain}
\section{Introduction}
Ultrasound measurements play very important roles in a $^3$He
investigation$^{1}$. The time scale of ultrasound corresponds to
quasiparticle life times at low temperatures and first to zero
sound crossover was observed. Energy scales of ultrasound match
the binding energies of Cooper pairs in superfluid $^3$He were
observed coupling through density oscillations. Sound transmission
method using many types of surface acoustic wave (SAW) sensors
have been developed for acoustical and electrical property
measurements of adjacent liquid or gas (especially liquid $^4$He
at low temperature$^{2}$). A Rayleigh-SAW propagates along the
substrate surface by emitting compressional waves into the quantum
fluid and thus the sampling of the Rayleigh-SAW is determined by
the acoustic impedance of the surrounding quantum fluids. Relevant
problems are the instability for the flow induced by the
SAW$^{3,4}$ or the slip effect existing along the interface of the
quantum fluid. In fact, the instability may triggers the
laminar-turbulent transition for flows of the superfluid. The
important issue is to determine the critical velocity$^{5,6}$ for
the relevant physical parameters so that the experimental
measurements could be under control.
\newline
A most striking characteristic of liquid helium is that it exists
in the liquid state down to the absolute zero temperature because
(i) the van der Waals forces in helium are weak; (ii) the
zero-point energy, due to the light mass, is large. In fact, it is
called a quantum liquid due to these kinds of quantum effects and
is closely related to the Bose-Einstein condensation for He II.
The well-known properties of He II can be largely accounted for on
the basis of phenomenological two-fluid theory$^{7,8}$. One of the
basic assumptions of the two-fluid model is : He II consists of a
kind of mixture of two components, a normal component and a
superfluid component. The former has viscosity while the
superfluid can move without friction as long as certain velocity
limits are not exceeded.
\newline
One crucial issue for the related researches about He II is the
critical velocity$^{5,6,9}$ (when it flows through a capillary or
plane channel) which depends on the micro-channel size. Landau
attributed the existence of a critical velocity in He II to the
breakdown of the superfluid due to the creation of excitations (he
proposed that phonons and rotons are two types of excitations
which make up the normal fluid). If the velocity is less than the
critical value, there will be no dissipation or friction along the
boundary or interface for the flow. In plane channel or slab flows
of quantum fluids, to determine the critical velocity corresponds
to finding out the critical Reynolds number$^{5-6,10}$.
\newline The traditional starting point of an investigation of
hydrodynamic stability is eigenvalue analysis, which proceeds in :
(i) linearize about the laminar solution and then (ii) look for
unstable eigenvalues of the linearized problem. In much of the
literature of hydrodynamic stability$^{10}$, attention has been
restricted to 2D perturbations, in particular, the well-known
Orr-Sommerfeld equation is an eigenvalue equation for 2D
perturbations. For pressure-driven flow, with the no-slip boundary
conditions (which are normally valid in macro-channels),
eigenvalue analysis$^{11}$ predicts a critical Reynolds number
$Re_{cr} =$ 5772 at which instability should first occur, but in
the laboratory$^{12}$, transition to turbulence is observed at
$Re_{cr}$ as low as $\sim$ 1000.
\newline
As for quantum fluids or liquids, there always exists a slip
velocity along the static interface or confinement due to the
microscopically incomplete momentum exchange therein$^{13}$. The
distinction among various flow regimes can be obtained by
introducing the Knudsen number (which also characterizes the value
of the slip velocity${13}$), Kn=mfp$/L$, where mfp is the mean
free path and $L$ is the characteristic flow dimension. Slip flow
conditions exist for $0.001<$Kn$ \le 0.1$, where the flow can be
considered continuous if the slip velocity at the walls are taken
into account.
\newline Meanwhile, the effect of elastic or deformable interfaces, like surface
acoustic waves (SAW) interacted with volume and surface phonons
(propagating along the elastic boundaries$^{14-17}$), upon the
stability of quantum fluid flows, however, were seldom considered
before, to the best knowledge of the authors. Although the
mathematical difficulty is essential therein. Note that,
entrainment of fluids induced by SAW propagating along deformable
boundaries have been studied$^{14}$ since early 1950s. The role of
elastic macroscopic walls resembles that of microscopic
phonons$^{4}$. As presented in [4], for the description of the
transport processes in the nonequilibrium gas-solid system
including the processes occurring in the case of propagation of
sound in a solid, we need to solve
\begin{displaymath}
 \frac{\partial f({\bf x},{\bf v},t)}{\partial t} +{\bf v}\cdot
 \frac{\partial f({\bf x},{\bf v},t)}{\partial {\bf x}}=I_g
 (\{f\}),
\end{displaymath}
\begin{displaymath}
 \frac{\partial n({\bf x},{\bf k}_j,t)}{\partial t} +{\bf c}_j\cdot
 \frac{\partial n({\bf x},{\bf k}_j,t)}{\partial {\bf x}}=I_v
 (\{n\}),
\end{displaymath}
\begin{displaymath}
 \frac{\partial H({\bf r},{\bf K}_{\xi},t)}{\partial t} +{\bf c}_{\xi}\cdot
 \frac{\partial H({\bf r},{\bf K}_{\xi},t)}{\partial {\bf r}}=I_s
 (\{f,n,H\}),
\end{displaymath}
and the associated boundary conditions (scattering and interacting
laws near the interface)
\begin{displaymath}
  |v_r| f^+ =\int_{v_i <0} d {\bf v}_i |v_i| f^- ({\bf v}_i)
  W({\bf v}_i \rightarrow {\bf v}_r),
\end{displaymath}
\begin{displaymath}
  \frac{|c_r|}{L_t} n^- ({\bf k}_j)=\sum_{{\bf k}_{j_1}(c_i >0)} [
  \frac{c_i}{L_t} n^+ ({\bf k}_{j_1})+\bar{N}_g ({\bf k}_{j_1})+
  \bar{N}_p ({\bf k}_{j_1})] V_p ({\bf k}_{j_1} \rightarrow {\bf
  k}_j; \omega).
\end{displaymath}
$f$, $n$, and $H$ denote the distribution function for gases,
volume phonons, and surface phonons, respectively. $I_g$, $I_v$,
and $I_s$ are the corresponding collision integrals. Please see
the details in [4] for other notations or symbols. The stability
problems for such a system of differential-integral equations
remain open up to now. To escape from above (many-body problems)
difficulties, we plan to use the macroscopic approach which is a
complicated extension of previous approaches$^{16,18}$.
\newline
In this work, the stability analysis of incompressible slip flows
for a normal fluid (above the transition temperature $T_c$) inside
a microchannel with the flow confined by two elastic (wall) layers
of thin films where surface (acoustic) waves are propagating along
the interfaces is conducted. Namely, we will relax the static- or
rigid-interface boundary conditions which are frequently used in
previous theoretical and/or experiment approaches into the
dynamic- or deformable-interface boundary conditions which are
more realistic in essence when we consider the flow stability
problem in a microdomain where surface acoustic waves$^{15}$
propagating along the boundaries of quantum fluids$^{2,4}$. The
verified code which was based on the spectral method developed by
Chu$^{18}$ will be extended here to include the boundary
conditions coming from SAW existing along the interfaces to obtain
the stability characteristics of the basic flow.
\newline This paper is organized as follows. We introduce the
mathematical formulation for the Orr-Sommerfeld equation and the
relevant linear stability analysis in Section 2. Boundary
conditions for the slip flow and the wavy interface of the fluid
system are then incorporated. The expression for primary slip
flows of quantum fluids we shall investigate their stability
characteristic will be derived before we describe the numerical
approach : a modified spectral method at the end of this Section.
Two physical parameters, $K_0$ (relevant to the SAW effect) and Kn
will be introduced or defined. Finally we shall present our
calculations and discuss them. Our results illustrate that the
critical Reynolds number ($Re_{cr}$) decreases (to 1441) rapidly
once the interfaces are subjected to propagating waves (or noises)
and there are slip velocities existing along the boundaries or
interfaces ($K_0 =1$ and Kn=0.001). However, the slip velocity
(adjusted by Kn) effect is minor and adverse compared to the SAW
effect (tuned by $K_0$) considering the decrease of $Re_{cr}$.
\section{Governing Equations}
Macroscopically, the motion of the normal fluid (above the
critical phase transition temperature of the quantum fluids) as a
whole could be treated  by using hydrodynamical models  starting
from the microscopic atomic wave function$^{8,19}$. Here, after
the simplifying treatment of the complicated mathematical
derivations, the dimensionless equations of motion for an
incompressible normal fluid flow$^{8,10,11,18}$, in the absence of
body forces and moments, reduce to
\begin{equation}
 \hspace*{2mm} \frac{\partial {\bf U}}{\partial t}+({\bf U} \cdot \nabla)
  {\bf U}=-\nabla P+\frac{1}{Re} \nabla^2 {\bf U}
\end{equation}
where ${\bf U}$ and $P$ stand for the velocity and pressure of
fluids. $Re =\rho u_{max} h/\mu$ is the Reynolds number with
$\rho$ and $\mu$ being the density and viscosity of fluids. For
the case of normal fluid flows driven by a constant
pressure-gradient, i.e., plane Poiseuille flow, the length scale
is the half width of the normal-fluid layer $h$, and the velocity
is the center-line velocity $u_{max}$. Following the usual
assumptions of linearized stability theory$^{10-11}$,
$U_{i}(x_{i},t)=\bar{u_{i}}(x_{i})+u'_{i}(x_{i},t)$, and
similarly, \hspace*{1mm} $P(x_{i},t)=\bar{p}(x_{i})+p'(x_{i},t)$,
the linearized equation, which governs the disturbances are:
\begin{equation}
 \hspace*{2mm} \frac{\partial u'_{i}}{\partial t}+(\bar{{\bf u}}\cdot
 \nabla)u_{i}'+({\bf u'}\cdot \nabla) \bar u_{i}=-\nabla p'+\frac{1}{Re}
 \nabla^2 u'_{i}
\end{equation}
Disregarding the lateral disturbances, $w'$=0, a stream function for the
disturbance, $\psi$, can be defined such that $u'=\partial \psi /
\partial y , v' =-\partial \psi / \partial x$. Using normal mode
decomposition analysis, $\psi$ may be assumed to have the form
$\psi(x,y,t)= \phi(y) \exp[i \alpha (x-Ct)]$, $\alpha$ is the wave
number (real), $C$ is $C_{r}+i C_{i}$. This is a kind of
Tollmien-Schlichting transversal waves, $C_{r}$ is the ratio
between the velocity of propagation of the wave of perturbation
and the characteristic velocity, $C_{i}$ is called the
amplification factor, and $\alpha$ equals to 2$\pi
{\Lambda}^{-1}$, where $\Lambda$ is the wave length of the
Tollmien-Schlichting perturbation$^{10}$. Substituting the stream
function and eliminating the pressure, we have the linearized
disturbance equation
\begin{equation}
 \hspace*{2mm}  (D^2-\alpha^2) (D^2-\alpha^2) \phi
 = i \alpha Re [(\bar{u} -C)(D^2-\alpha^2)\phi-(D^2 \bar{u})\phi ]
\end{equation}
where $D=d/dy$. This is also valid for the slip flow
regime$^{13}$, $0.001<K_n \le 0.1$, since the flow can still be
considered as continuous.
\subsection{Boundary Conditions}
For the slip flow, the continuous models can be used if the
no-slip boundary condition is modified. A few models have been
suggested to estimate the nonzero velocity at a boundary
surface$^{13,16}$. In this study, we adopt the approach based on
Taylor's expansion of the velocity around the wall. Thus, the
first order approximation yields $\bar u \mid_{wall}=\mbox{Kn} d
\bar u /dy$ (positive for the inner normal as $y\equiv n$).
Consequently, the mean (basic) velocity profile is given by
\vspace*{0.5mm}
\begin{equation}
  \bar{u}=1-y^2+2 \mbox{Kn}
\end{equation}
for $-1\le y \le 1$. Boundary conditions for $\phi$ or $D \phi$
are not the same as previous no-slip approach, i.e., $\phi(\pm
1)=D \phi (\pm 1)=0$ and shall be introduced below.
\subsection{Interface Treatment}
We consider a two-dimensional layer (slab) of uniform thickness
filled with a homogeneous normal fluid (Newtonian viscous fluid;
its dynamics is  described by Navier-Stokes equations). The upper
and lower boundaries of the layer are superfluids which are rather
flexible, on which are imposed travelling sinusoidal waves of
small amplitude $a$ (due to SAW or peristaltic waves). The
vertical displacements of the upper and lower interfaces ($y=h$
and $-h$) are thus presumed to be $\eta$ and $-\eta$,
respectively, where $\eta=a \cos \frac{2\pi}{\lambda} (x-ct)$,
$\lambda$ is the wave length, and $c$ the wave speed. $x$ and $y$
are Cartesian coordinates, with $x$ measured in the direction of
wave propagation and $y$ measured in the direction normal to the
mean position of the interfaces.
It would be expedient to simplify these equations by
introducing dimensionless variables. We have a characteristic
velocity $c$ and three characteristic lengths $a$, $\lambda$, and
$h$. The following variables based on $c$ and $h$ could thus be
introduced :
\begin{displaymath}
 x'=\frac{x}{h}, \hspace*{3mm} y'=\frac{y}{h}, \hspace*{6mm}
 u'=\frac{u}{c}, \hspace*{3mm} v'=\frac{v}{c}, \hspace*{2mm}
 \eta'=\frac{\eta}{h}, \hspace*{3mm} \psi'=\frac{\psi}{c\,h}, \hspace*{4mm}
 t'=\frac{c\,t}{h}, \hspace*{3mm} p'=\frac{p}{\rho c^2},
\end{displaymath}
where $\psi$ is the dimensional stream function. The amplitude
ratio $\epsilon$, the wave number $\alpha$, and the Reynolds
number $Re_c$ are defined by
\begin{displaymath}
 \epsilon=\frac{a}{h}, \hspace*{4mm} \alpha=\frac{2 \pi h}{\lambda},
 \hspace*{4mm} Re_c =\frac{c\,h}{\nu}.
\end{displaymath}
After introducing the dimensionless variables, now, for the ease
and direct representation of our mathematical expressions in the
following, we shall drop those primes (') in those dimensionless
variables and treat them as dimensionless.  We seek a solution in
the form of a series in the parameter $\epsilon$ :
\begin{displaymath}
 \psi=\psi_0 +\epsilon \psi_1 + \epsilon^2 \psi_2 + \cdots,
 \hspace*{12mm} \frac{\partial p}{\partial x}=(\frac{\partial p}{\partial
 x})_0+\epsilon (\frac{\partial p}{\partial x})_1 +\epsilon^2 (\frac{\partial
 p}{\partial x})_2 +\cdots,
\end{displaymath}
with $u=\partial \psi/\partial y$, $v=-\partial \psi/\partial x$.
The 2D (x- and y-) momentum equations and the equation of
continuity for the normal fluid could be in terms of the stream
function $\psi$ if the pressure ($p$) term is eliminated. The
final governing equation is
\begin{equation}
 \frac{\partial}{\partial t} \nabla^2 \psi + \psi_y \nabla^2 \psi_x
 -\psi_x \nabla^2 \psi_y =\frac{1}{Re_c}\nabla^4 \psi,
\hspace*{12mm} \nabla^2 \equiv\frac{\partial^2}{\partial x^2}
+\frac{\partial^2}{\partial y^2}   ,
\end{equation}
and subscripts indicate the partial differentiation. If we presume
originally the fluid is quiescent; this corresponds to a free
pumping case and finally the velocity profile of the fluid is
symmetric with respect to the centerline of the plane channel
bounded by the superfluids. Equation above, together with the
condition of symmetry and a uniform constant pressure-gradient in
the x-direction, $(\partial p/\partial x)_0$=constant, yield :
\begin{equation}
 \psi_0 =K_0 [ (1+2 \mbox{Kn}) y-\frac{y^3}{3}],  \hspace*{24mm}
 K_0=\frac{Re_c}{2}(-\frac{\partial p}{\partial x})_0 , 
\end{equation}
$K_0$ is in fact a necessary pumping to sustain a plane Poiseuille
flow (pressure-driven case). $\psi_0$  corresponds to the solution
of
\begin{equation}
 \frac{\partial}{\partial t} \nabla^2 \psi_0 +\psi_{0y} \nabla^2
 \psi_{0x}-\psi_{0x} \nabla^2 \psi_{0y}=\frac{1}{Re_c} \nabla^4
 \psi_0 ,
\end{equation}
and
\begin{equation}
 \psi_1 =\frac{1}{2}\{\phi(y) e^{i \alpha (x-t)}+\phi^* (y) e^{-i \alpha
 (x-t)} \} , 
\end{equation}
where the asterisk denotes the complex conjugate.
The normal fluid is subjected to boundary conditions imposed by
the symmetric motion of the wavy interfaces and the slip condition
at interfaces. The basic slip flow now has this form (cf [16]), as
$u=\partial \psi_0/\partial y$,
\begin{equation}
 \bar{u}= 1-y^2+2 \mbox{Kn},
\end{equation}
where $c$ is the phase speed of the SAW, Kn=mfp/$h$. Boundary
conditions become
\begin{equation}
 \phi_y (\pm 1) \pm\phi_{yy} (\pm 1) \mbox{Kn}=2 K_0 (1\pm \mbox{Kn}),
 \hspace*{24mm}
 \phi (\pm 1)=\pm 1.
\end{equation}
\subsection{Numerical Approach}   
The eigenvalue problem raised above could be solved by using the
verified code$^{18}$, which used the spectral method$^{20}$ based
on the Chebyshev-polynomial-expansion approach, once the equation
and boundary conditions are discretized. For instance, we have,
from equation (3), as a finite-sum approximation (reduction from
$\infty$ to N),
\begin{displaymath}
 \phi (z)=\sum^N_{n=0} a_n T_n (z),
\end{displaymath}
where $T_n (z)$ is the Chebyshev polynomial$^{11,20}$ of degree
$n$ with $z=\cos (\theta)$. $T_n (z)$ are known to satisfy the
recurrence relations
\begin{displaymath}
 z T_n (z)=\frac{1}{2}[T_{n+1} (z)+ T_{n-1} (z)].
\end{displaymath}
After substituting $\phi$ into (3) and with tremendous
manipulations, we obtain the algebraic equation
$$  \frac{1}{24} \sum^N_{\stackrel{\scriptstyle p=n+4}{p\equiv
   n(\mbox{\small mod} \,2)}}
    [ p^{3} (p^{2}-4)^{2}-3n^{2}p^{5}+     \\
  3n^{4}p^{3} -p n^{2}(n^{2}-4)^{2} ] a_{p}-         $$
$$  \sum^N_{\stackrel{\scriptstyle p=n+2}{p\equiv n(\mbox{\small mod}
    \,2)}} \{ [2\alpha^{2}+ \\
  \frac{1}{4} i \alpha Re(4\,M_0 f-4 C -M_0 c_{n}-M_0 c_{n-1})] p(p^{2}-n^{2})- \\
  \frac{1}{4} i \alpha Re M_0 c_{n}p[p^{2}-(n+2)^{2}] -  $$
$$ \frac{1}{4} i \alpha Re M_0 d_{n-2}p[p^{2}- (n-2)^{2}] \} a_{p}+ i \alpha \\
   Re M_0 n(n-1)a_{n}+ \{\alpha^{4}+i \alpha Re[(M_0 f-C)\alpha^{2}-2 M_0] \}  \\
   c_{n}a_{n} -   $$
\begin{equation}
 \hspace*{1mm} \frac{1}{4}i \alpha^{3} Re M_0 [c_{n-2}a_{n-2}+ c_{n}(c_{n}+
 c_{n-1})a_{n} + c_{n}a_{n+2}]=0
\end{equation}
for $n\geq 0$, $f=1+2 K_{n}$, where $c_{n}=0$ if $n > 0$, and
$d_{n}=0$ if $n < 0$, $d_{n}=1$ if n$\geq$ 0. Here, $M_0 =1$, $C$
is the complex eigenvalue.  The boundary conditions become
\begin{equation}
 \sum^{N}_{\stackrel{\scriptstyle n=1}{n \equiv 1(\mbox{\small mod} \,2)}}
 a_{n} = 1, \hspace*{6mm} \\
 \sum^{N}_{\stackrel{\scriptstyle n=1}{n \equiv 1(\mbox{\small mod} \,2)}}
 [n^{2}+\mbox{Kn}\frac{n^2 (n^2-1)}{3}] a_{n} = 2 K_0.
\end{equation}
The matrices thus formed are of poor condition because they are
not diagonal, symmetric$^{21}$. Thus, before we perform
floating-point computations to get the complex eigenvalues, we
precondition these complex matrices to get less errors. Here we
adapt Osborne's algorithm to precondition these complex matrices
via rescaling, i.e., by certain diagonal similarity
transformations of the matrix (errors are in terms of the
Euclidean norm of the matrix) designed to reduce its norm. The
details of this algorithm could be traced in [18,21-22]. The form
of the reduced matrix is upper Hessenberg. Finally we perform the
stabilized $LR$ transformations for these matrices to get the
complex eigenvalues (please see also [18,22] for the details). The
preliminary verified results of this numerical code  had been done
for the cases of Kn=$0$ (no-slip boundary conditions) in
comparison with the bench-mark results of Orszag's$^{11}$ . For
example, for $Re=10000.0$, $\alpha=1.0$ of the test case : plane
Poiseuille flow, we obtained the same spectra as $0.23752648+$ i
$0.00373967$ for $C_r+$ i $C_i$ which Orszag obtained in 1971.
\section{Results and Discussion}
After careful verification, we proceed to obtain (through
tremendous searching using double-precision machine accuracy) the
detailed spectra for the illustration of the stability of the slip
flow in normal fluids confined between wavy interfaces. To
demonstrate some of the calculated spectra ($C_r, C_i$) near the
regime of ($Re_{cr},\alpha$), we plot Fig. 1 by selecting 2 pairs
of ($Re, \alpha$)=$(1562, 1.156)$ and $(2982.3, 1.0783)$ with the
corresponding $K_0$= 1, and 0.5 for the same Knudsen number
(Kn=0.01). Once $C_i >0$, the instability occurs! The onset of
instability is easy to occur once the Reynolds number, wave
number, or Knudsen number perturb a little again near this regime
($C_i$ becomes zero and then positive). \newline We then plot the
neutral stability boundary curves for different cases in Fig. 2.
It is clear that each curve is composed of two branches (one is
upper and the other is lower, and they coalesce into a critical
point ($Re_{cr}$ and $\alpha_{cr}$) as the Reynolds number
decreases). We tune the $K_0$ parameter to be 1 and 0.5, with the
corresponding Knudsen number being $0.001$ and $0.01$,
respectively. Once the Knudsen number is set to be zero and there
is no SAW effect, we recover the curve obtained by Orszag$^{11}$
($Re_{cr}\sim 5772$). Otherwise, the resulting critical Reynolds
numbers ($Re_{cr}$) are 1441,1562, 2664, 2982.3, respectively.  It
seems the effect of SAW propagating along the interface is the
dominated one and will degrade the flow stability significantly.
The slip velocity effect is minor and adverse (delay the
transition).
\newline To understand the stability behavior related to the decay or
amplification of the perturbed disturbance waves in the finite
time for certain mode, we also illustrate their time evolution
patterns by selecting the least unstable mode. As illustrated in
Figs. 3 and 4 for Re=1441, $\alpha=1.175$ ($K_0=1$) and Re=2664,
$\alpha=1.105$ ($K_0=0.5$), we can observe the oscillating or
amplifying pattern just after a finite time (time is dimensionless
and the Knudsen number is the same, Kn=0.001). The original
disturbance (wave) will not decay for these unstable modes ($C_r,
C_i$)$\sim(0.382,0.000002)$, and $\sim(0.324,0.00000046)$,
respectively.
\newline We can finally conclude that various kinds of interface noises
(as illustraed here, slip velocities and the propagating surface
acoustic wavs) will premature any instability mechanism
considering the temporal growth of the disturbances. We have
obtained more clues about the slip flow (which is in a
non-equilibrium state) instability of quantum fluids (above their
critical transition temperature) by considering more realistic
interface conditions. Once we know the viscosities and/or
densities of these quantum fluids, based on the obtained critical
Reynolds number, we can then determine the critical velocity$^{
5,6}$ for each case. Meanwhile, these results will help
researchers to understand the formation or generation of vorticity
waves and then the route to low-temperature turbulence in quantum
fluids. It seems the range of wave numbers relevant to the SAW
propagating along the flexible interfaces, the Knudsen numbers and
the Reynolds numbers of basic slip flows of normal fluids must be
carefully selected for the optimal flow control usage in SAW
applications$^{1,2}$ to the investigation of $^3$He. Our further
study will be relevant to those more complicated issues
$^{2-3,22-25}$.
\newline {\small Acknowledgements. The author is partially supported by the National
Natural Science Foundation of China (NNSFC) under grant No. :
10274061 and the China Post-Dr. Science Foundation (grant No.
:1999-17). }

\newpage


\psfig{file=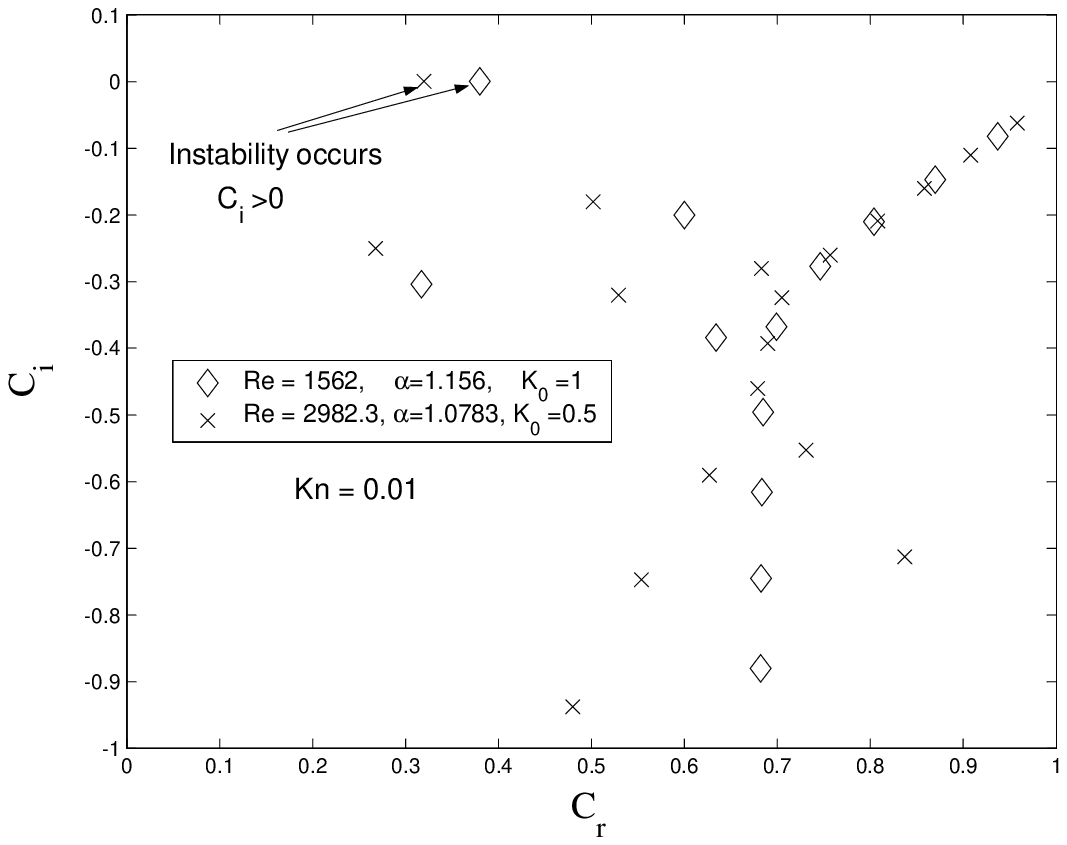,bbllx=0.0cm,bblly=10.6cm,bburx=12cm,bbury=23.8cm,rheight=9.6cm,rwidth=9.6cm,clip=}
%

\begin{figure}[h]
\hspace*{12mm} Fig. 2 \hspace*{2mm}  Illustration of the temporal
spectra ($C_r, C_i$) for disturbance \newline \hspace*{12mm} waves
due to interface ($K_0$) and slip velocity (Kn=$0.01$) effects.
\newline \hspace*{12mm} $Re = 1562, 2982.3$ for corresponding
$K_0=1, 0.5$  and  $\alpha=1.156, 1.0783$, respectively.
\end{figure}

\psfig{file=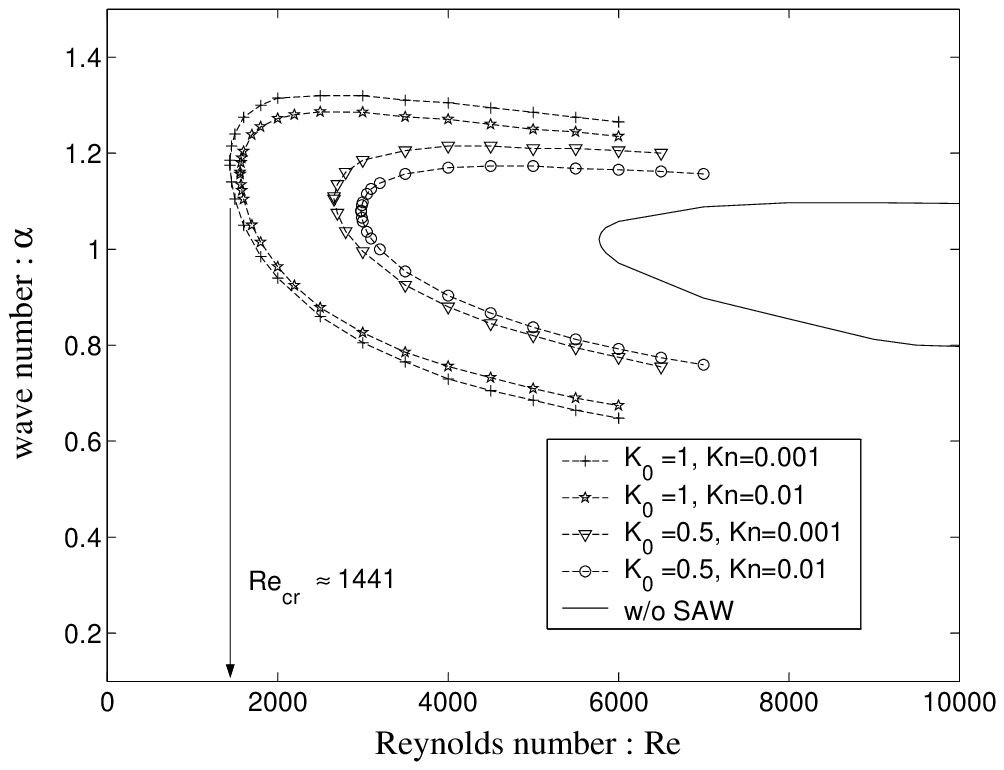,bbllx=0.0cm,bblly=12.5cm,bburx=12cm,bbury=24.8cm,rheight=9.6cm,rwidth=9.6cm,clip=}

\begin{figure}[h]
\hspace*{12mm} Fig. 3 \hspace*{2mm} Comparison of  wavy interface
($K_0$) and slip velocity (Kn) effects
\newline \hspace*{12mm} on the neutral stability boundary of the basic
flow. Kn= mfp/$h$. mfp is  \newline \hspace*{12mm}
 the mean free path of the quantum fluid.  $Re_{cr} \sim 1441, 1562, 2664, 2982.3$
\newline \hspace*{12mm}for $K_0=1$ : Kn=$0.001, 0.01$, and $K_0=0.5$ : Kn=$0.001, 0.01$.
\newline \hspace*{12mm}
\end{figure}

\newpage

\psfig{file=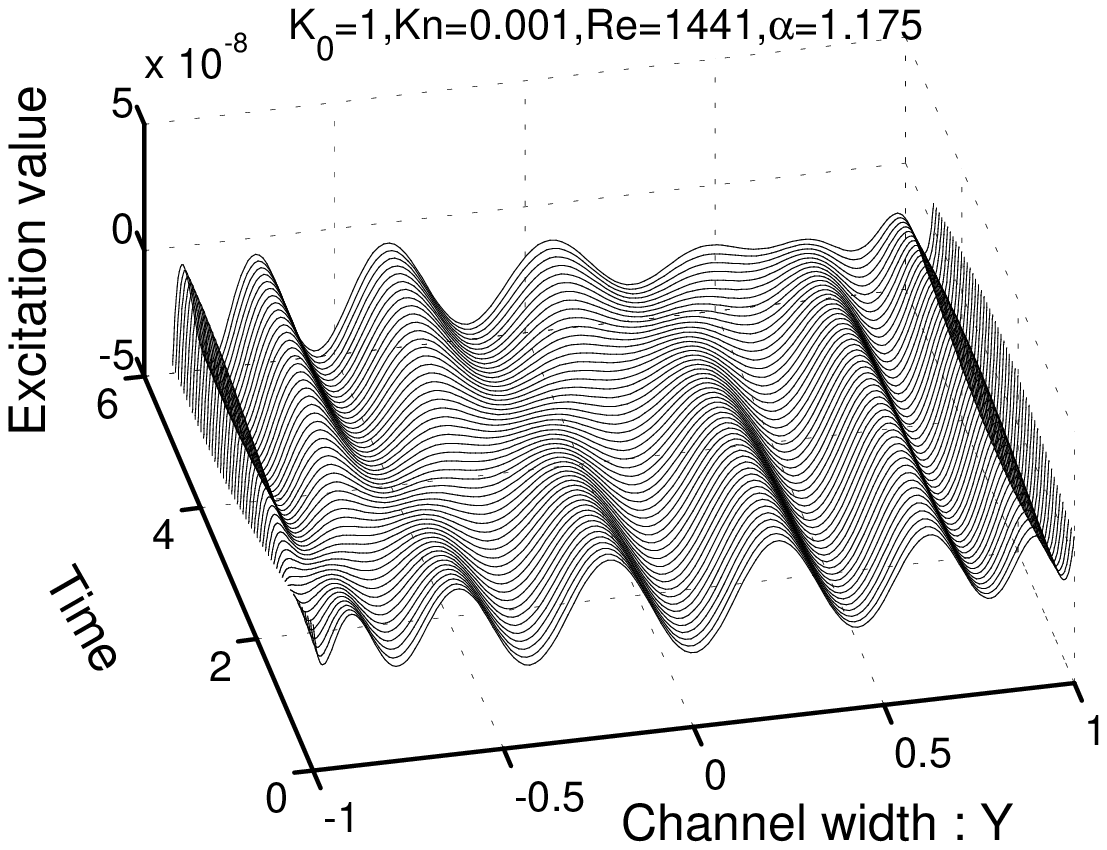,bbllx=0.2cm,bblly=11cm,bburx=15cm,bbury=24.4cm,rheight=9.8cm,rwidth=9.8cm,clip=}
\begin{figure}[h]

\hspace*{12mm} Fig. 3 \hspace*{2mm}  Illustration of the temporal
evolution for disturbance \newline \hspace*{12mm} waves due to
interface ($K_0$) and slip velocity (Kn) effects.
\newline \hspace*{12mm} $Re = 1441$ for corresponding
$K_0$=1  and  $\alpha$=1.175, Kn=0.001. Time is dimensionless.

\end{figure}

\psfig{file=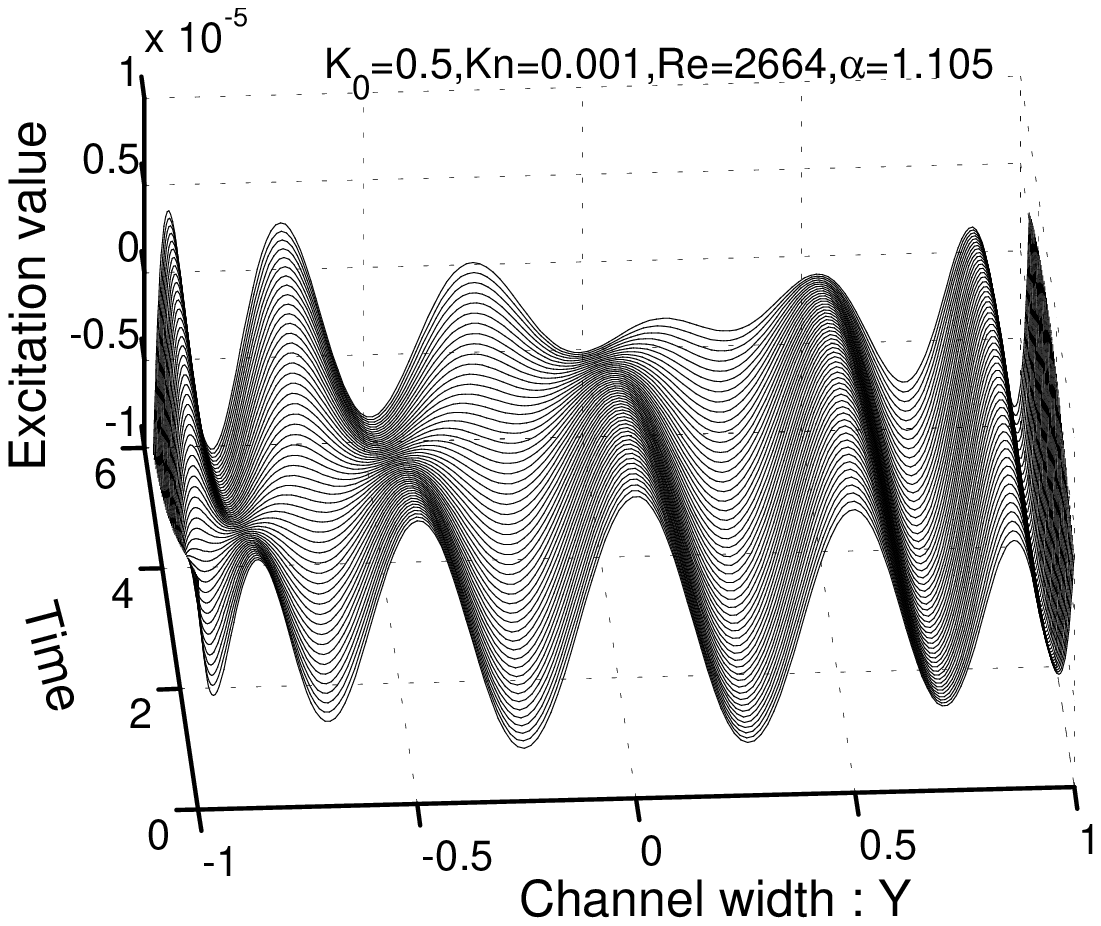,bbllx=0.2cm,bblly=11cm,bburx=15cm,bbury=24.4cm,rheight=9.8cm,rwidth=9.8cm,clip=}
\begin{figure}[h]

\hspace*{12mm} Fig. 4 \hspace*{2mm} Illustration of the temporal
evolution for disturbance \newline \hspace*{12mm} waves due to
interface ($K_0$) and slip velocity (Kn) effects.
\newline \hspace*{12mm} $Re = 2664$ for corresponding
$K_0$=0.5  and  $\alpha$=1.105, Kn=0.001. Time is dimensionless.
\end{figure}


\end{document}